\documentclass[%
 reprint,
 amsmath,amssymb,
 aps,
]{revtex4-1}

\usepackage{graphicx}% Include figure files
\usepackage{dcolumn}% Align table columns on decimal point
\usepackage{bm}% bold math
\usepackage{epstopdf}
\begin{document}

\preprint{APS/123-QED}

\title{Deviation from canonical collective creep behavior in Li$_{0.8}$Fe$_{0.2}$OHFeSe}% Force line breaks with \\
%\thanks{A footnote to the article title}%

\author{Yue Sun,$^{1*}$ Sunseng Pyon$^1$, Run Yang,$^2$ Xianggang Qiu,$^2$ Jiajia Feng,$^3$ Zhixiang Shi,$^3$ and Tsuyoshi Tamegai$^1$}

\affiliation{%
$^1$Department of Applied Physics, The University of Tokyo, Tokyo 113-8656, Japan \\
$^2$Beijing National Laboratory for Condensed Matter Physics, Institute of Physics, Chinese Academy of Sciences, Beijing 100190, People's Republic of China\\
$^3$Department of Physics and Key Laboratory of MEMS of the Ministry of Education, Southeast University, Nanjing 211189, China}

\date{\today}% It is always \today, today,
             %  but any date may be explicitly specified

\begin{abstract}
We present an extensive study of the vortex dynamics in Li$_{0.8}$Fe$_{0.2}$OHFeSe single crystal, which is prepared by intercalating insulating spacer layer Li$_{0.8}$Fe$_{0.2}$OH into FeSe. Absence of fishtail effect and the crossover from elastic to plastic creep are observed similar to the parent compound FeSe. However, when we apply collective creep theory to the magnetic relaxation data, the obtained creep exponent $\mu$ is found to be $\sim$ 4.1 much larger than the predicted maximum value. Besides, the elastic creep in the vortex phase diagram of Li$_{0.8}$Fe$_{0.2}$OHFeSe is found to reside only in an extremely small region. Such a large value of $\mu$ and the small elastic creep region may be originated from the weakening of coupling between the vortices in neighbouring layers by elongating the layer distance. It may indicate that the vortex structure of  Li$_{0.8}$Fe$_{0.2}$OHFeSe is in the crossover regime between elastic Abrikosov vortices to stacks of pancake vortices. Our study also suggests that the intercalated FeSe is a promising candidate for searching intrinsic Josephson phenomena.

%\begin{description}
%\item[PACS numbers]
%\verb+74.25.Wx+, \verb+74.25.Uv+, \verb+74.25.Sv+, \verb+74.70.Xa+

%\end{description}
\end{abstract}

%\pacs{Valid PACS appear here}% PACS, the Physics and Astronomy
                             % Classification Scheme.
%\keywords{Suggested keywords}%Use showkeys class option if keyword
                              %display desired
\maketitle
\section{introduction}
Since the discovery of high-temperature cuprate superconductors, the vortex structure and its relation with the layered crystal structure became a crucial issue directly related to both the microscopic physics as well as its potential for applications. When the $c$-axis coherence length, $\xi_c$, is larger than the layer distance, $d$, such as the YBa$_2$Cu$_3$O$_{7-\delta}$ (YBCO), the Abrikosov vortices of flux lines will be formed via interlayer coupling, which will collectively creep together when the thermal activation is weaker than the pinning energy \cite{Blatterreview,Yeshurunreview}. On the other hand, the interlayer coupling is very weak when $\xi_c < d$ such as in Bi$_2$Sr$_2$CaCu$_2$O$_{8+y}$ (BSCCO), and hence the vortices can be considered as stacks of pancake vortices, which can creep individually in each layer \cite{Blatterreview,Yeshurunreview}. The former case is directly related to the high-power application since it limits the current-carrying capacity of superconductors \cite{Kwokreview}. On the other hand, the latter is essential for realizing intrinsic Josephson junctions \cite{THzgapnatpho}.

In these two cases, the vortex dynamics, pinning behavior, and vortex phase diagram are all quite different \cite{Blatterreview}. Thus, it is interesting to know the behavior of vortices in between these two extreme situations. The evolution of vortex physics with crystal structure is also crucial for the application research, and instructive for searching for and designing new structures. However, there are only a few materials \cite{MollPRLiJJ} or artificial layered structures in the crossover regime \cite{UstinovPRB}. Besides, it is also difficult to compare the vortex physics of samples from different systems since their band structure, paring mechanism, and other fundamental properties are different. To solve these issues, it is ideal to study the vortex physics in the same system with different interlayer distance. The intercalated FeSe is such a good candidate. The elastic Abrikosov vortices have been reported in the parent FeSe \cite{SunPhysRevBJcFeSe} because of the relatively short interlayer distance ($\sim$ 5.5 ${\textrm{\AA}}$) and the small anisotropy ($\gamma$ $\sim$ 2) \cite{TerashimaPRB}. On the other hand, the interlayer distance can be enhanced by intercalating spacer layers \cite{BurrardNatMat,Lunatmat,Guonatcomm,SakaiPRB,HosonoJPSJ}, and it can be gradually tuned up to $\sim$ 19 ${\textrm{\AA}}$ by choosing different kinds of intercalated layers \cite{HosonoJPSJ,HatakedaJPSJ}. At the same time, the superconducting transition temperature, $T_c$, is also enhanced with increasing the interlayer distance \cite{HosonoJPSJ}, which is advantageous for applications.

In this report, we focused on the material of Li$_{0.8}$Fe$_{0.2}$OHFeSe (prepared by intercalating insulating spacing layer Li$_{0.8}$Fe$_{0.2}$OH into the FeSe) because the large-size single crystals with high quality can be easily obtained \cite{DongxiaoliPhysRevB.92.064515}. Through extensive studies on vortex dynamics in this system, an anomalously large value of vortex creep exponent $\mu$ $\sim$ 4.1, which exceeds the maximum value based on the collective creep theory, has been extracted. Besides, the elastic creep is found to reside only in an extremely small region of the vortex phase diagram. These results indicate that the interlayer coupling in FeSe is weakened by elongating the interlayer distance, and the vortex structure of Li$_{0.8}$Fe$_{0.2}$OHFeSe is in the crossover regime between elastic Abrikosov vortices and stacks of pancake vortices. Our preliminary study also suggests that the intercalated FeSe is a promising candidate for realizing intrinsic Josephson junctions.

\section{experiments}
Li$_{0.8}$Fe$_{0.2}$OHFeSe single crystals were grown by the hydrothermal method as described in Ref. \cite{Lunatmat,DongxiaoliPhysRevB.92.064515}. Large crystals of K$_{0.8}$Fe$_{1.6}$Se$_2$ were firstly synthesized as the matrix. Then, 2 g single crystals of K$_{0.8}$Fe$_{1.6}$Se$_2$ mixed with 0.02 mole selenourea, 0.0075 mole Fe powder, and 12 g LiOH$\cdot$H$_2$O were put into a Teflon-lined steel autoclave (25 ml) mixed with 10 ml de-ionized water. After heating at 160 $^\circ$C for 72 hours. The K ions in K$_{0.8}$Fe$_{1.6}$Se$_2$ were completely released into solution after the hydrothermal reaction process and shining single crystals of (Li$_{1-x}$Fe$_x$)OHFeSe along (00$l$) direction were left. Magnetization measurements were performed using a commercial SQUID magnetometer (MPMS-XL5, Quantum Design). The magnetic relaxation rate, $S$ = $\mid$dln$M$/dln$t$$\mid$, was measured by tracing the decaying of magnetization with time $M(t)$ due to creep motion of vortices for more than one hour, where $t$ is the time from the moment when the critical state is prepared. In these measurements, magnetic field was swept more than 5 kOe higher than the target field to ensure the full penetration of the field.

\section{results and discussion}
\begin{figure}\center
\includegraphics[width=8.5cm]{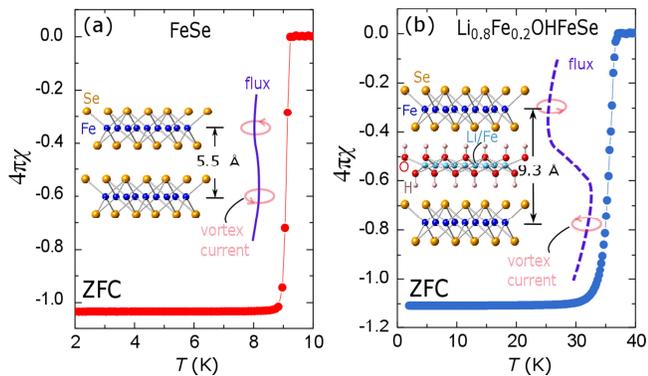}\\
\caption{(Color online) Temperature dependencies of ZFC magnetizations for (a) FeSe and (b) Li$_{0.8}$Fe$_{0.2}$OHFeSe measured at 5 Oe for $H \parallel c$. Insets are the crystal structures of FeSe and Li$_{0.8}$Fe$_{0.2}$OHFeSe}\label{}
\end{figure}

Insets of Figs. 1(a) and (b) show the crystal structures of FeSe and Li$_{0.8}$Fe$_{0.2}$OHFeSe, respectively. It is obvious that the distance between neighbouring FeSe layers are enhanced from 5.5 ${\textrm{\AA}}$ in FeSe to 9.3 ${\textrm{\AA}}$ in Li$_{0.8}$Fe$_{0.2}$OHFeSe after intercalating the spacer layer \cite{Lunatmat}. The main panels of Figs. 1(a) and (b) compare the temperature dependencies of zero-field-cooled (ZFC) magnetization at 5 Oe for the two crystals. The value of $T_{\rm{c}}$ is enhanced from $\sim$ 9 K to  $\sim$ 36 K by inserting Li$_{0.8}$Fe$_{0.2}$OH layer between FeSe layers, which is similar to the previous report \cite{Lunatmat}.

\begin{figure}\center
\includegraphics[width=8cm]{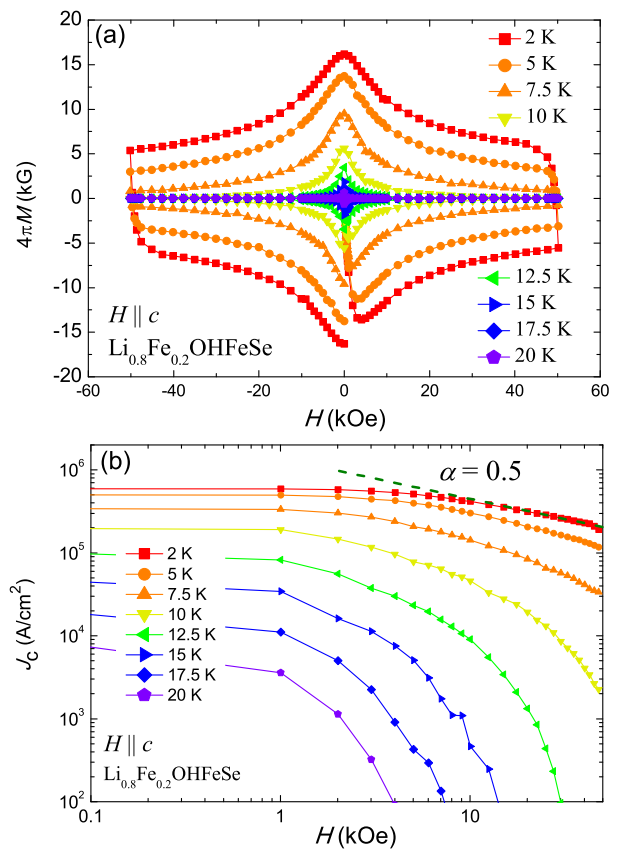}\\
\caption{(Color online) (a) Magnetic hysteresis loops of Li$_{0.8}$Fe$_{0.2}$OHFeSe single crystal at different temperatures ranging from 2 to 20 K for $H$ $\parallel$ $c$. (b) Magnetic field dependence of critical current densities for $H$ $\parallel$ $c$.}\label{}
\end{figure}

Figure 2(a) shows the magnetic hysteresis loops (MHLs) of Li$_{0.8}$Fe$_{0.2}$OHFeSe obtained at different temperatures for \emph{H} $\|$ \emph{c}. The symmetric loops indicate that the bulk pinning is dominating in this crystal. The value of $M$ is monotonically decreasing with increasing $H$, namely, the second magnetization peak (also known as fish-tail effect), is absent. Such a monotonic MHL is similar to that of the parent FeSe single crystal \cite{SunPhysRevBJcFeSe}, while it is different from fish-tail effects observed in most other iron-based superconductors (IBSs) including Fe(Te,Se) \cite{ProzorovPRBBaCoJc,HaberkornPRBCaNa,ShenBingPRB,SalemSuguiPRB,SunAPEX,SunEPL,TaenFeTeSePRB}. One possible explanation of the fish-tail effect is the competition between strong and weak pinnings \cite{vanderBeekPRL2010,SunPhysRevBJcFeSe}. The strong pinning is attributed to sparse nanometer-sized defects naturally formed during the crystal growth \cite{vanderBeekPRB2002}. The weak pinning is attributed to the atomic-scaled defects, which are mainly from the electron/hole doping induced charged quasi-particle-scattering centers, such as oxygen vacancies in \textit{RE}FeAsO$_{1-x}$ (\textit{RE}: rare earth), electron doping Co atom in Ba(Fe$_{0.93}$Co$_{0.07}$)$_2$As$_2$ \cite{FeigelmanPRL,Blatterreview}. The absence of fish-tail effect in FeSe is explained by the dominance of strong pinnings since it is innately superconducting without electron/hole doping induced charged quasi-particle-scattering centers \cite{SunPhysRevBJcFeSe}. The absence of fish-tail effect is sustained after intercalating spacer layer, suggesting that the pinning mechanism is not changed, i.e. strong-pinning is still dominant in Li$_{0.8}$Fe$_{0.2}$OHFeSe.

The absence of fish-tail effect can be witnessed more clearly in the field dependence of critical current density, $J_{\rm{c}}$, in Fig. 2(b) obtained by using the extended Bean model:\cite{Beanmodel}
\begin{equation}
\label{eq.1}
J_{\rm{c}}=20\frac{\Delta M}{a(1-a/3b)},
\end{equation}
where $\Delta$\emph{M} is \emph{M}$_{\rm{down}}$ - \emph{M}$_{\rm{up}}$, \emph{M}$_{\rm{up}}$ [emu/cm$^3$] and \emph{M}$_{\rm{down}}$ [emu/cm$^3$] are the magnetization when sweeping fields up and down, respectively, \emph{a} [cm] and \emph{b} [cm] are sample widths (\emph{a} $<$ \emph{b}). The self-field $J_{\rm{c}}$ reaches a value $\sim$6 $\times$ 10$^5$ A/cm$^2$ at 2 K, which is more than one order larger than that of FeSe single crystal \cite{SunPhysRevBJcFeSe}. The value of $J_{\rm{c}}$ is similar to that reported in high-quality FeTe$_{1-x}$Se$_x$ single crystals prepared by annealing in controlled atmosphere \cite{TaenFeTeSePRB,SunAPEX,SunSciRep,*SunSUSTFeSeannealing,*SunJPSJTeannealing,*YamadaJPSJannealing,*ChenJingtingJPSJannealing}. The $J_{\rm{c}}$ changes little below 1 kOe, followed by a power-law decay $H^{-\alpha}$ with $\alpha$ $\sim$ 0.5 at low temperatures. Such behavior is also observed in most IBSs as well as in FeSe, which is attributed to the strong point pinnings by sparse nm-sized defects \cite{vanderBeekPRB2010,SunPhysRevBJcFeSe,vanderBeekPRB2002}. At higher temperatures, the $J_{\rm{c}}$ decreases faster with field, which is caused by the large vortex creep rate, and will be discussed later.

\begin{figure}\center
\includegraphics[width=8cm]{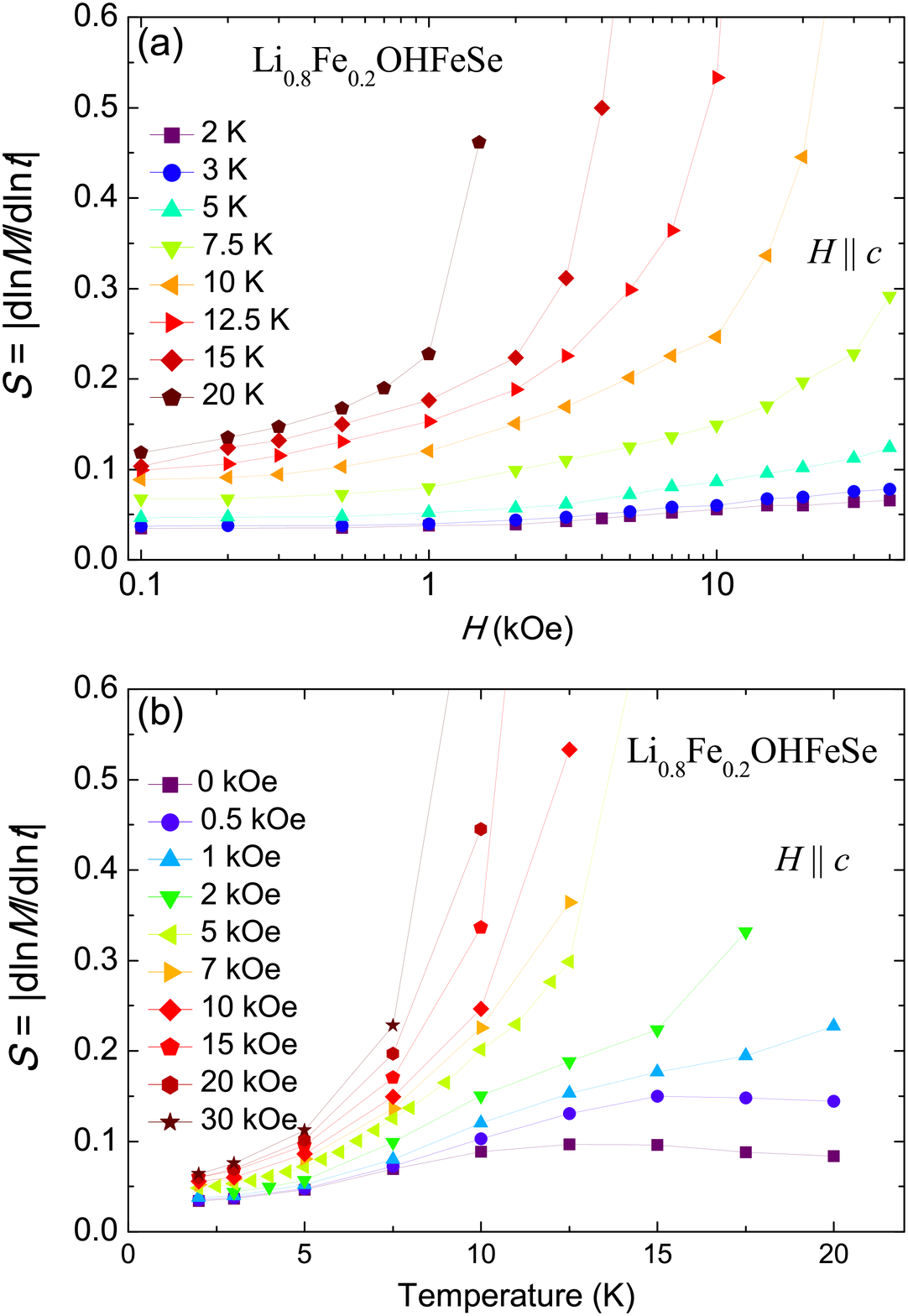}\\
\caption{(Color online) (a) Temperature dependence of normalized magnetic relaxation rate $S$ at different fields. (b) Magnetic field dependence of $S$ at different temperatures.}\label{}
\end{figure}

Figure 3(a) shows the field dependence of normalized magnetic relaxation rate $S$ at temperatures ranging from 2 to 20 K. Under small fields, $S$ keeps a relatively small value, while it slightly increases with field. When the field is larger than a certain temperature-dependent value, $S$ starts to increase much faster with field. A similar monotonic increase of $S$ with field is also observed in FeSe \cite{SunPhysRevBJcFeSe}, which is different from YBa$_2$Cu$_3$O$_{7-\emph{$\delta$}}$ \cite{CivalePRB}, ``122"-type IBSs \cite{ProzorovPRBBaCoJc,NakajimaPRBirra}, and FeTe$_{1-x}$Se$_x$ \cite{SunEPL}, where $S$ shows a steep decrease with field at low fields. Such a steep decrease in $S$  is usually attributed to crossover from single vortex regime at low fields with a small creep exponent $\mu$ = 1/7 to small or large bundle regime at higher fields with larger $\mu$. The absence of non-monotonic field dependence of $S$ indicates that the single vortex regime may not exist or only exist at temperatures lower than the measurement limit of 2 K. The sudden increase in $S$ with field is also observed in the $S - T$ curves as shown in Fig. 3(b). Before the steep increase, a temperature insensitive plateau is observed at low temperatures under small fields with a relatively large vortex creep rate (e.g. 0.09 $<$ $S$ $<$ 0.15 at $\sim$15 K below 0.5 kOe). Such plateau-like behavior of $S$ has been observed in FeSe, and other high temperature superconductors \cite{Yeshurunreview,EleyNatMat,ProzorovPRBBaCoJc,HaberkornPRBBaCo,HaberkornPRBCaNa,ShenBingPRB,TaenPRBBaCoirra,SunEPL,YangHPRB}, which is usually interpreted by the collective creep theory \cite{Yeshurunreview}.

\begin{figure}\center
\includegraphics[width=8cm]{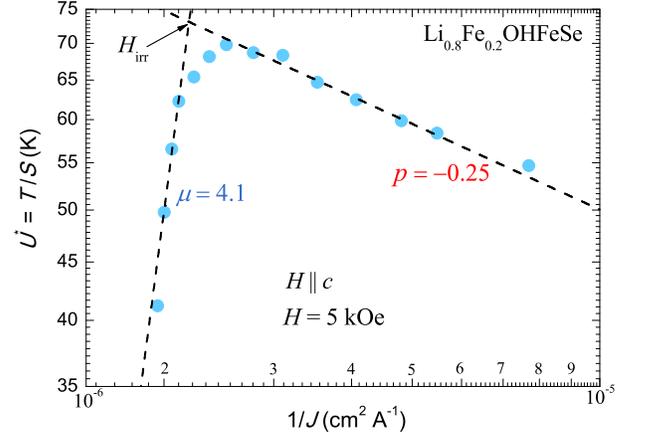}\\
\caption{(Color online) Inverse current density dependence of effective pinning energy \emph{U}$^*$ at 5 kOe in Li$_{0.8}$Fe$_{0.2}$OHFeSe single crystal. The dashed lines show the fitting results based on Eq. (5).}\label{}
\end{figure}

In the collective creep theory \cite{Yeshurunreview}, magnetic relaxation rate $S$ can be given by
\begin{equation}
\label{eq.2}
S=\frac{T}{U_0+\mu Tln(t/t_{\rm{eff}})},
\end{equation}
where $U_0$ is the flux activation energy in the absence of flux creep, \emph{t}$_{\rm{eff}}$ is the effective hopping attempt time, and $\mu$ is the exponent for collective creep. The value of $\mu$ contains information about the size of the vortex bundle in the collective creep theory. In a three-dimensional system, it is predicted as $\mu$ = 1/7, (1) 5/2, 7/9 for single-vortex, (intermediate) small-bundle, and large-bundle regimes, respectively \cite{Blatterreview,FeigelmanPRL}. The flux activation energy \emph{U} as a function of current density \emph{J} can be defined as \cite{FeigelmanPRB}
\begin{equation}
\label{eq.3}
U(J)=\frac{U_0}{\mu}[(J_{\rm{c0}}/J)^\mu-1].
\end{equation}
Combining this with $U$ = $T$ln($t$/$t_{\rm{eff}}$) extracted from the Arrhenius relation, we can deduce the so-called interpolation formula
\begin{equation}
\label{eq.4}
J(T,t)=\frac{J_{\rm{c0}}}{[1+(\mu T/U_0)ln(t/t_{\rm{eff}})]^{1/\mu}},
\end{equation}
where $J_{\rm{c0}}$ is the temperature dependent critical current density in the absence of flux creep. From Eqs. (3) and (4), effective pinning energy $U$$^*$ = $T$/$S$ can be derived as
\begin{equation}
\label{eq.5}
U^*=U_0+\mu Tln(t/t_{\rm{eff}})=U_0(J_{\rm{c0}}/J)^\mu.
\end{equation}

Based on the equation above, we estimated the value of $\mu$ from the slope in the double logarithmic plot of $U$$^*$ vs 1/$J$. Fig. 4(a) shows a typical result at 5 kOe. The first data point at 2 K is slightly away from the straight line. This deviation may be originated from the effect of quantum creep, which usually enhances at low temperature, and hence reduces the value of $U^\ast$.  The evaluated value of $\mu$ is $\sim$ 4.1, which is larger than the predicted maximum value based on the collective creep theory. A similar large value of $\mu$ is also reported in Na-doped CaFe$_2$As$_2$ single crystal, which is explained by the large contribution from quantum creep \cite{HaberkornPRBCaNa}. However, the estimated quantum creep rate in Li$_{0.8}$Fe$_{0.2}$OHFeSe is only $\sim$ 0.018 ($S_q \simeq \frac{e^2\rho_{\rm{n}}}{\hbar \xi}(\frac{J_{\rm{c}}}{J_0})^2$ \cite{QuantumcollectivecreepPhysRevLett.66.3297}, where
$\rho_{\rm{n}}\simeq$ 150 $\mu\Omega$cm is the normal-state resistivity, $\xi\simeq$ 2 nm is coherence length \cite{ChunleiWangsustgamma}. $J_0=c\phi_0/12\sqrt3\pi^2\xi_{\rm{ab}}\lambda_{\rm{ab}}^2$ is depairing current density, which is estimated as 6$\times$10$^7$ A/cm$^2$ with the penetration depth $\lambda_{\rm{ab}}\simeq$ 283 nm \cite{KhasanovPhysRevB.93.224512}.), much smaller than the experimental value at low temperatures, which is too small to affect the value of $\mu$. Origin of such a large $\mu$ will be discussed in detail later. Contrary to the above prediction of $\mu$ $>$ 0, a negative slope with value $\sim$ -0.25 is obtained at small $J$. The negative slope is often denoted as $p$ in the plastic creep theory, which is thought to lead to faster escape of vortices \cite{AbulafiaPRL}. These results indicate that the crossover, which is usually treated as elastic to plastic creep, is also observed in Li$_{0.8}$Fe$_{0.2}$OHFeSe similar to FeSe \cite{SunPhysRevBJcFeSe}.

\begin{figure}\center
\includegraphics[width=8cm]{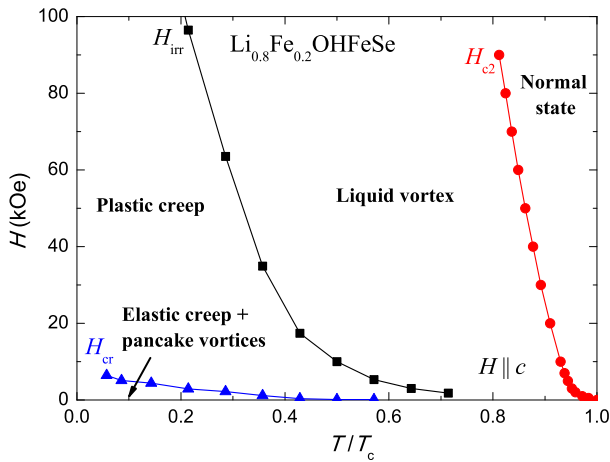}\\
\caption{(Color online) The vortex phase diagram for Li$_{0.8}$Fe$_{0.2}$OHFeSe single crystal with $H$ $\|$ $c$.}\label{}
\end{figure}

With the crossover obtained under different fields and temperatures, we constructed a vortex phase diagram for Li$_{0.8}$Fe$_{0.2}$OHFeSe single crystal as shown in Fig. 5. $H_{\rm{c2}}$ represents the upper critical field obtained from the midpoint of the resistive transition at $T_c$ under fields up to 90 kOe. $H_{\rm{irr}}$ is the irreversibility field obtained by extrapolating $J_{\rm{c}}$ to zero in $J_c^{1/2}$ vs $H$ curves. $H_{\rm{cr}}$ is the crossover field. At fields lower than $H_{\rm{cr}}$, vortices in Li$_{0.8}$Fe$_{0.2}$OHFeSe creep elastically at a relatively low speed. With increasing field over $H_{\rm{cr}}$, the vortex creep rate increases, and vortices start to move via plastic deformation. Upon further increase in field over $H_{\rm{irr}}$, vortex pinning becomes no longer effective and vortex solid transforms into unpinned vortex liquid. Such a phase diagram is similar to that of FeSe and other IBSs. However, it is very unique in Li$_{0.8}$Fe$_{0.2}$OHFeSe that the elastic creep only exists in an extremely small region in the $H-T$ phase diagram. When temperature is increased over 50\% of $T_{\rm{c}}$ or the field is increased to $\sim$ 8 kOe (Considering the $H_{\rm{c2}}$ of Li$_{0.8}$Fe$_{0.2}$OHFeSe is $\sim$ 800 kOe for $H \parallel c$ \cite{DongxiaoliPhysRevB.92.064515}, it is only 1\% of the $H_{\rm{c2}}$), the vortex system enters fast-moving plastic creep region. It indicates that the vortex pinning  in Li$_{0.8}$Fe$_{0.2}$OHFeSe is much weaker than that in FeSe, which is consistent with the large value of  $\mu$ obtained above.

Now, we turn to the discussion about the origin of the unexpectedly large $\mu$ and the extremely small elastic creep region. In FeSe and most IBSs, the anisotropy is small ($\gamma$ $\sim$ 1.5-3) due to the small distance between the neighbouring FeSe/FeAs layers compared to the coherence length $\xi_c$. In such a case, the coupling between the vortices in neighboring layers is strong enough to form well-defined Abrikosov vortices. When the thermal energy is smaller than the pinning energy, the Abrikosov vortices will collectively creep in the form of single vortex line, small-bundle, or large-bundle depending on the driving force and the pinning situation. On the other hand, in superconductors with large anisotropy such as BSCCO ($\gamma$ $>$ 100), the interlayer coupling is weak because of the large separation between superconducting layers, and vortices consist of stacks of 2D pancake vortices that are coupled mainly via magnetic coupling. In this situation, pancake vortices can creep individually in each layer, without collective pinning. In Li$_{0.8}$Fe$_{0.2}$OHFeSe, the interlayer distance is largely elongated compared to that in FeSe as shown in the insets of Figs. 1(a) and (b), and the anisotropy also increases to larger than 10 \cite{DongxiaoliPhysRevB.92.064515,ChunleiWangsustgamma}, which result in weak coupling between vortices in neighboring layers. Compared to the extreme case of BSCCO, the Abrikosov vortices may be still formed in Li$_{0.8}$Fe$_{0.2}$OHFeSe since the slowly increasing region of $S$ with $T$, as well as the crossover from elastic to plastic creeps are still observed. However, the coupling is very weak so that the vortices can also individually creep to some extent within the layer as shown by the sketch in the insets of Figs. 1(a) and (b). Such in-plane creep of individual vortices together with the collective creep of the Abrikosov vortices increase the creep rate of vortices, and causes apparent increase of $\mu$ when we force the collective creep theory to fit the relaxation data. Besides, the in-plane creep may make the Abrikosov vortices unstable, and hence relatively small thermal activation overcomes the collective pinning, and activates the vortices to creep with a larger rate into the plastic creep region. It can explain the extremely small elastic creep region in the vortex phase diagram. Thus, the vortex structure of Li$_{0.8}$Fe$_{0.2}$OHFeSe is found to be in the crossover regime between elastic Abrikosov vortices and stacks of pancake vortices. Our results also indicate that systematic study on a series of intercalated FeSe with different interlayer separations is promising to reveal the evolution of vortex physics with crystal structure.

On the other hand, periodic entry of Josephson vortices, which can connect neighboring pancake vortices, has been observed in IBSs (V$_2$Sr$_4$O$_6$)Fe$_2$As$_2$ with a large interlayer distance \cite{MollNatPhys}. Such results proved that intrinsic Josephson junctions can be also realized in multi-band superconductors, which offers another way to study intrinsic Josephson junctions in system other than BSCCO \cite{KleineriJJBSCCOPhysRevLett.68.2394}. However, its complicated crystal structure makes the growth of large single crystal or thin film of (V$_2$Sr$_4$O$_6$)Fe$_2$As$_2$ with acceptable quality very difficult, and hinders the study of intrinsic Josephson junctions in IBSs. On the other hand, large single crystals of Li$_{0.8}$Fe$_{0.2}$OHFeSe with good quality have been reported to grow easily \cite{DongxiaoliPhysRevB.92.064515}, and recently the successful growth of thin film has also been reported \cite{LiFeOHFeSethinfilm}. Thus, our current research suggests that the intercalated FeSe system is a promising candidate for searching and applying intrinsic Josephson phenomena of IBSs. Recently, the layer distance in intercalated FeSe has been enhanced over 19 ${\textrm{\AA}}$ \cite{HosonoJPSJ,HatakedaJPSJ}, in which the intrinsic Josephson junction is highly anticipated.

\section{conclusions}
In summary, we have studied the vortex dynamics of Li$_{0.8}$Fe$_{0.2}$OHFeSe with $T_{\rm{c}}$ $\sim$ 36 K and compared it with the parent FeSe by performing magnetization measurements to deduce critical current density and magnetic relaxation rate. The vortex creep rate with an unexpected large exponent $\mu$ $\sim$ 4.1 is obtained. The crossover from elastic to plastic creep are observed similar to those in FeSe.  However, the elastic creep region in the vortex phase diagram of Li$_{0.8}$Fe$_{0.2}$OHFeSe is found to reside only in an extremely small region. Such large value of vortex creep exponent $\mu$ and the extremely small elastic creep region indicate that the vortex structure of  Li$_{0.8}$Fe$_{0.2}$OHFeSe is in the crossover regime of elastic Abrikosov vortices to stacks of pancake vortices. Our preliminary study suggests that the intercalated FeSe is a promising candidate for hosting intrinsic Josephson phenomena.

\acknowledgements
The present work was partly supported by KAKENHI (Grants No.18H05853 and No.17H01141) and a bilateral project between Japan and China supported by JSPS. R.Y. and X.Q. thank the support from MOST of China (973 Projects No. 2017YFA0302900) and NSFC (Grants No. 11374345, No.91421304). J.F. and Z.S. thank the support from NSFC (Grants No. 11611140101).

$^{*}$Present address: Department of Physics and Mathematics, Aoyama Gakuin University, Sagamihara 252-5258, Japan.
E-mail: sunyue@phys.aoyama.ac.jp 

\bibliography{references}

\end{document}